\documentclass[a4paper,12pt]{article}          
\usepackage{graphicx}
\usepackage{amssymb}
\usepackage{slashbox}
\usepackage{amsmath}
\usepackage{amsbsy}
\usepackage{dsfont}
\begin{document}

\title{A Complete Transient Analysis for the Incremental LMS Algorithm}

\author{Muhammad Omer Bin Saeed}


\date{}

\maketitle

\begin{abstract}
The incremental least mean square (ILMS) algorithm was presented in \cite{Lopes2007}. The article included theoretical analysis of the algorithm along with simulation results under different scenarios. However, the transient analysis was left incomplete. This work presents the complete transient analysis, including the learning behavior. The analysis results are verified through several experimental results.
\end{abstract}


\section{Introduction}
\label{intro}

Over the last decade, several algorithms have been proposed for distributed estimation over wireless sensor networks. Different algorithms target different goals. However, the most significant feature among these algorithms has been the basic distribution schemes used for developing the algorithms. Nearly all algorithms have been based on either the incremental scheme or the diffusion scheme \cite{Sayed2014}.

The incremental scheme was first introduced in \cite{Lopes2007}. The network was distributed using a Hamiltonian cycle and each node was connected to two other nodes in the cycle. The resulting algorithm converged quicker than the diffusion scheme and gave lower misadjustment as well. While the diffusion scheme presented some advantages over the incremental scheme, one significant advantage was the theoretical analysis \cite{Lopes2008}. Since the analysis of the diffusion scheme is easier to perform, most subsequent algorithms have been based on the diffusion scheme \cite{Sayed2014}.

The analysis of the incremental scheme is not easy to perform. The analysis presented in \cite{Lopes2007} is incomplete. The transient analysis presented in \cite{Lopes2007} reaches a certain point before the focus shifts to steady-state results. The reason is that the learning behavior for the incremental scheme is not straightforward. As a result, the subsequent algorithms based on the ILMS algorithm of \cite{Lopes2007} also follow the same process and leave the transient analysis incomplete. This work presents the complete transient analysis for the ILMS algorithm, including the learning behavior. The results are verified through several experimental results. The experiments are conducted to test all aspects of the analysis and simulation results are used to validate the theoretical findings.

The rest of the paper is divided as follows. Section \ref{algo} presents the ILMS algorithm. Section \ref{new_an} presents the mean square analysis in details, including the transient and steady-state analysis. Experimental results are presented in section \ref{results} and section \ref{conc} gives the conclusion.

\section{The ILMS Algorithm}
\label{algo}

\begin{figure}[h]
\centering{\includegraphics[width=75mm]{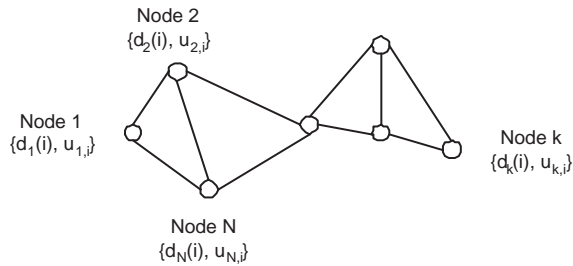}}
\caption{An illustration of an adaptive network of $N$ nodes.}\label{Fig1}
\end{figure}

A collection of $N$ sensor nodes, spread over a geographical area is being considered. The nodes are connected in a Hamiltonian cycle \cite{Lopes2007}. Fig. \ref{Fig1} shows an illustration of a possible adaptive wireless sensor network. Connecting the nodes in a cyclic way would result in the incremental strategy.

The unknown parameters are modeled as a vector, ${\bf w}_o$, of size ($M \times 1$). The input to a node at any given time instant, $i$, is a ($1 \times M$) regressor vector, ${\bf u}_k(i)$, where $k = 1,\hdots,N$ is the node index. The resulting observed output for the node is a noise corrupted scalar, $d_k(i)$, given by
\begin{equation}\label{eq1}
d_k(i) = {\bf u}_k(i){\bf w}^o + v_k(i),
\end{equation}
where $v_k(i)$ is the zero-mean additive noise with variance $\sigma^2_{v,k}$.

The incremental least mean square (ILMS) algorithm is given by \cite{Lopes2007}
\begin{eqnarray}
{\bf w}_k(i) &=& {\bf w}_{k-1} (i) + \mu_k e_k(i) {\bf u}_k^T (i),  \label{eq2} \\
{\bf w} (i) &=& {\bf w}_N (i), \label{eq3}
\end{eqnarray}
where ${\bf w}_k(i)$ is the estimate of the unknown vector at time instant $i$ for node $k$, $e_k(i) = d_k(i) - {\bf u}_k(i){\bf w}_{k-1} (i)$ is the instantaneous error, ${\bf w} (i)$ is the final estimate for iteration $i$ and $(.)^T$ is the transpose operator.

\section{Proposed analysis}
\label{new_an}

Let the weight-error vector be given by
\begin{equation}\label{eq4}
{\bf \tilde w}(i) = {\bf w}_o - {\bf w}(i).
\end{equation}
Using (\ref{eq4}) in (\ref{eq2}) and simplifying gives
\begin{eqnarray}
\nonumber{\bf \tilde w}_k(i) &=& [{\rm {\bf I}}_M - \mu_k {\bf u}_k^T(i){\bf u}_k(i)]{\bf \tilde w}_{k-1}(i) \\
&& - \mu_k {\bf u}_k^T(i) v_k(i), \label{eq5}
\end{eqnarray}
where ${\rm {\bf I}}_M$ is an identity matrix of size $M$. The auto correlation matrix of the input regressor vector is given by ${\bf R}_{{\bf u}_k} = {\mathds E}[{\bf u}_k^T(i){\bf u}_k(i)]$, where ${\mathds E}[.]$ is the expectation operator. ${\bf R}_{{\bf u}_k}$ can be decomposed into its component matrices of eigenvalues and eigenvectors. Thus, ${\bf R}_{{\bf u}_k} = {\bf H}_k {\bf \Lambda}_k {\bf H}_k^T$, where ${\bf H}_k$ is the matrix of eigenvectors such that ${\bf H}_k^T{\bf H}_k = {\rm {\bf I}}_M$ and ${\bf \Lambda}_k$ is a diagonal matrix containing the eigenvalues. Using the matrix ${\bf H}_k$, the following transformations are made
\[
\begin{array}{*{20}c}
{{\bf \overline w}_k(i)  = {\bf{H}}_k^T {\bf{\tilde w}}_k(i) }, & {{\bf{\overline u}}_k(i)  = {\bf{u}}_k(i) {\bf{H}}_k}  \\
\end{array}
\]
The weight-error update equation thus becomes
\begin{eqnarray}
\nonumber {\bf \overline w}_k(i) &=& [{\rm {\bf I}}_M - \mu_k {\bf \overline u}_k^T(i){\bf \overline u}_k(i)]{\bf \overline w}_{k-1}(i) \\
&&- \mu_k {\bf \overline u}_k^T(i) v_k(i). \label{eq6}
\end{eqnarray}

\subsection{Mean Analysis}
\label{mean_an}

Applying the expectation operator to (\ref{eq6}) gives
\begin{eqnarray}
\nonumber {\mathds E}\left[{\bf \overline w}_k(i)\right] &&= {\mathds E}\left[ \left\{{\rm {\bf I}}_M - \mu_k {\bf \overline u}_k^T(i){\bf \overline u}_k(i) \right\}{\bf \overline w}_{k-1}(i)\right. \\
\nonumber && \left. \hspace{0.5cm} - \mu_k {\bf \overline u}_k^T(i) v_k(i) \right] \\
\nonumber &&= \left\{ {\rm {\bf I}}_M - \mu_k {\mathds E} \left[ {\bf \overline u}_k^T(i){\bf \overline u}_k(i) \right] \right\} {\mathds E} \left[ {\bf \overline w}_{k-1}(i) \right] \\
&&= \left\{ {\rm {\bf I}}_M - \mu_k {\bf \Lambda}_k \right\} {\mathds E} \left[ {\bf \overline w}_{k-1}(i) \right], \label{eq7}
\end{eqnarray}
where the data independence assumption separates ${\mathds E}[{\bf w}(i)]$ from the rest of the variables. The second term is 0 as additive noise is independent and zero-mean and ${\mathds E} \left[ {\bf \overline u}_k^T(i){\bf \overline u}_k(i) \right] = {\bf \Lambda}_k$. The sufficient condition for stability is evaluated from (\ref{eq7}) and is given by
\begin{equation}\label{eq8}
0 < \mu_k < \frac{2}{{\beta _{k,max} } },
\end{equation}
where ${\beta_{k,max}}$ is the maximum eigenvalue of ${\bf \Lambda}_k$.

\subsection{Mean-Square Analysis}
\label{mean_sq_an}

For the mean-square analysis, the approach of \cite{Lopes2007} is followed. However, the analysis in \cite{Lopes2007} was not completed to give the learning behavior for the algorithm. Here, the learning behavior is also given in closed form. Taking the squared weighted $l_2$-norm of (\ref{eq6}) and applying the expectation operator yields
\begin{eqnarray}\label{eq9}
&&{\mathds E}\left[ \left\| {\bf \overline w}_k (i) \right\|^2 _{{\bf \Sigma}} \right] \hspace{0cm} \\
\nonumber &&= {\mathds E}\left[ {\bf \overline w}_{k-1}^T(i) {\bf \Sigma}_k^{'} {\bf \overline w}_{k-1}(i) \right] + {\mathds E} \left[ \mu_k^2 v_k^2 (i) {\bf \overline u}_k(i) {\bf \Sigma}_k {\bf \overline u}_k^T (i) \right]\\
\nonumber && - {\mathds E} \left[ \mu_k v_k (i) {\bf \overline u}_k (i) {\bf \Sigma}_k \left\{{\rm {\bf I}}_M - \mu_k {\bf \overline u}_k^T (i) {\bf \overline u}_k (i) \right\} {\bf \overline w}_{k-1}(i) \right] \\
\nonumber && - {\mathds E} \left[ {\bf \overline w}_{k-1}^T(i) \left\{{\rm {\bf I}}_M - \mu_k {\bf \overline u}_k^T (i) {\bf \overline u}_k (i) \right\} {\bf \Sigma}_k \mu_k v (i) {\bf \overline u}_k^T (i) \right],
\end{eqnarray}
where $\left\|.\right\|$ is the $l_2$-norm operator and ${\bf \Sigma}_k$ is a weighting matrix. The weighting matrix ${\bf \Sigma}_k^{'}$ is given by
\begin{eqnarray}\label{eq10}
\nonumber {\bf \Sigma}_k^{'} &=& \left\{{\rm {\bf I}}_M - \mu_k {\bf \overline u}_k^T (i) {\bf \overline u}_k (i) \right\}^T {\bf \Sigma}_k \left\{{\rm {\bf I}}_M - \mu_k {\bf \overline u}_k^T (i) {\bf \overline u}_k (i) \right\} \\
\nonumber &=& {\rm {\bf I}}_M - \mu_k {\bf \overline u}_k^T (i) {\bf \overline u}_k (i) {\bf \Sigma}_k - \mu_k {\bf \Sigma}_k {\bf \overline u}_k^T (i) {\bf \overline u}_k (i) \\
&&+ \mu_k^2 {\bf \overline u}_k^T (i) {\bf \overline u}_k (i) {\bf \Sigma}_k {\bf \overline u}_k^T (i) {\bf \overline u}_k (i)
\end{eqnarray}
The last two terms in (\ref{eq9}) are zero since the additive noise is independent. Using the data independence assumption, the remaining two terms are simplified as
\begin{eqnarray}\label{eq11}
{\mathds E}\left[ \left\| {\bf \overline w}_k (i) \right\|^2 _{{\bf \Sigma}_k} \right] &=& {\mathds E}\left[ \left\| {\bf \overline w}_{k-1} (i) \right\|^2 _{{\bf \Sigma}_k^{'}} \right] \\
\nonumber &&+ \sigma^2_{v,k} \mu_k^2 {\rm Tr} \left\{ {\bf \Sigma}_k {\bf \Lambda}_k \right\},
\end{eqnarray}
where $\sigma^2_{v,k}$ is the additive noise variance at node $k$, ${\rm Tr} \left\{.\right\}$ is the {\em trace} operator and ${\mathds E} \left[ {\bf \overline u}_k(i) {\bf \Sigma}_k {\bf \overline u}_k^T (i) \right] = {\rm Tr} \left\{ {\bf \Sigma}_k {\bf \Lambda}_k \right\}$.

The expectation operator from the first term in (\ref{eq11}) applies to the weighting matrix independently as well since the data is being assumed to be independent \cite{Sayedbook}. Thus, we have, after simplification
\begin{eqnarray}\label{eq12}
\nonumber {\bf \Sigma}_k^{'} &=& {\rm {\bf I}}_M - 2 \mu_k {\bf \Lambda}_k {\bf \Sigma}_k + \mu_k^2 {\bf \Lambda}_k {\rm Tr} \left[ {\bf \Sigma}_k {\bf \Lambda}_k \right] \\
&&+ \mu_k^2 {\bf \Lambda}_k {\bf \Sigma}_k {\bf \Lambda}_k.
\end{eqnarray}
Using the $\mbox {diag}\{.\}$ operator, (\ref{eq11}) is further simplified to
\begin{eqnarray}\label{eq13}
{\mathds E}\left[ \left\| {\bf \overline w}_k(i) \right\|^2 _{{\boldsymbol \sigma}_k} \right] &=& {\mathds E}\left[ \left\| {\bf \overline w}_{k-1} (i) \right\|^2 _{{\bf F}_k {\boldsymbol \sigma}_k} \right] \\
\nonumber &&+ \sigma^2_{v,k} \mu_k^2 {\boldsymbol \lambda}_k^T {\boldsymbol \sigma}_k
\end{eqnarray}
where ${\boldsymbol \sigma}_k = {\mbox{diag}} \left\{ {\bf \Sigma}_k \right\}$, ${\boldsymbol \lambda}_k = {\mbox{diag}}\{{\bf \Lambda}_k\}$ and ${\mbox{diag}} \left\{ {\bf \Sigma}_k^{'} \right\} = {\boldsymbol \sigma}_k^{'} = {\bf F}_k {\boldsymbol \sigma}_k$, where ${\bf F}_k$ is given by
\begin{equation}\label{eq14}
{\bf F}_k = {\rm {\bf I}}_M - 2 \mu_k {\bf \Lambda}_k + \mu_k^2 \left[{\bf \Lambda}_k^2  + {\boldsymbol \lambda}_k {\boldsymbol \lambda}_k^T\right].
\end{equation}

Now, it can be seen from (\ref{eq14}) that ${\bf F}_k$ remains fixed at every iteration, even if it varies for each node, depending on the individual values of the parameters. Thus, using (\ref{eq13}), the analysis is initialized as
\begin{equation*}
{\mathds E} \left[ \left\| {\bf \overline w}(0) \right\|^2_{{\boldsymbol \sigma}} \right] = \left\| {\bf w}_o \right\|^2_{{\boldsymbol \sigma}}, \\
\end{equation*}
The first iterative update for node $1$ is given by
\begin{eqnarray*}
{\mathds E} \left[ \left\| {\bf \overline w}_1 (1) \right\|^2_{{\boldsymbol \sigma}_1} \right] &&= {\mathds E} \left[ \left\| {\bf \overline w}(0) \right\|^2_{{\bf F}_1{\boldsymbol \sigma}_1} \right] + \sigma^2_{v,1} \mu_1^2 {\boldsymbol \lambda}_1^T {\boldsymbol \sigma}_1 \\
&&= \left\| {\bf w}_o \right\|^2_{{\bf F}_1{\boldsymbol \sigma}_1} + \sigma^2_{v,1} \mu_1^2 {\boldsymbol \lambda}_1^T {\boldsymbol \sigma}_1
\end{eqnarray*}
Similarly, the first update for node $2$ is given by
\begin{eqnarray*}
{\mathds E} \left[ \left\| {\bf \overline w}_2(1) \right\|^2_{{\boldsymbol \sigma}_2} \right] &=& {\mathds E} \left[ \left\| {\bf \overline w}_1(1) \right\|^2_{{\bf F}_2{\boldsymbol \sigma}_2} \right] + \sigma^2_{v,2} \mu_2^2 {\boldsymbol \lambda}_2^T {\boldsymbol \sigma}_2 \\
&=& \left\| {\bf w}_o \right\|^2_{{\bf F}_1{\bf F}_2{\boldsymbol \sigma}_2} + \sigma^2_{v,1} \mu_1^2 {\boldsymbol \lambda}_1^T {\bf F}_2 {\boldsymbol \sigma}_2 \\
&&+ \sigma^2_{v,2} \mu_2^2 {\boldsymbol \lambda}_2^T {\boldsymbol \sigma}_2
\end{eqnarray*}
Continuing for node $3$ gives
\begin{eqnarray*}
{\mathds E} \left[ \left\| {\bf \overline w}_3 (1) \right\|^2_{{\boldsymbol \sigma}_3} \right] &=& {\mathds E} \left[ \left\| {\bf \overline w}_2 (1) \right\|^2_{{\bf F}_3 {\boldsymbol \sigma}_3} \right] + \sigma^2_{v,3} \mu_3^2 {\boldsymbol \lambda}_3^T {\boldsymbol \sigma}_3 \\
&=& \left\| {\bf w}_o \right\|^2_{{\bf F}_1 {\bf F}_2 {\bf F}_3 {\boldsymbol \sigma}_3} + \sigma^2_{v,1} \mu_1^2 {\boldsymbol \lambda}_1^T {\bf F}_2 {\bf F}_3 {\boldsymbol \sigma}_3 \\
&&+ \sigma^2_{v,2} \mu_2^2 {\boldsymbol \lambda}_2^T {\bf F}_3 {\boldsymbol \sigma}_3 + \sigma^2_{v,3} \mu_3^2 {\boldsymbol \lambda}_3^T {\boldsymbol \sigma}_3 \\
&=& \left\| {\bf w}_o \right\|^2_{{\bf \overline F}_3 {\boldsymbol \sigma}_3} + {\bf B}_3 {\boldsymbol \sigma}_3,
\end{eqnarray*}
where
\begin{eqnarray*}
{\bf \overline F}_3 &=& {\bf F}_1 {\bf F}_2 {\bf F}_3 = \prod\limits_{m=1}^{3} {{\bf F}_m} \\
{\bf B}_3 &=& \sigma^2_{v,1} \mu_1^2 {\boldsymbol \lambda}_1^T {\bf F}_2 {\bf F}_3 + \sigma^2_{v,2} \mu_2^2 {\boldsymbol \lambda}_2^T {\bf F}_3 + \sigma^2_{v,3} \mu_3^2 {\boldsymbol \lambda}_3^T {\bf I}_M \\
&=& \sum\limits_{m=1}^{2} {\sigma^2_{v,m} \mu_m^2 {\boldsymbol \lambda}_m^T \left\{ \prod\limits_{n=m+1}^{3} {{\bf F}_n} \right\} } + \sigma^2_{v,3} \mu_3^2 {\boldsymbol \lambda}_3^T {\bf I}_M.
\end{eqnarray*}
Thus, for node $k$, the first update is
\begin{equation*}
{\mathds E} \left[ \left\| {\bf \overline w}_k (1) \right\|^2_{{\boldsymbol \sigma}_k} \right] = \left\| {\bf w}_o \right\|^2_{{\bf \overline F}_k {\boldsymbol \sigma}_k} + {\bf B}_k {\boldsymbol \sigma}_k,
\end{equation*}
where
\begin{eqnarray*}
{\bf \overline F}_k &=& \prod\limits_{m=1}^{k} {{\bf F}_m} \\
{\bf B}_k &=& \sum\limits_{m=1}^{k-1} {\sigma^2_{v,m} \mu_m^2 {\boldsymbol \lambda}_m^T \left\{ \prod\limits_{n=m+1}^{k} {{\bf F}_n} \right\} } + \sigma^2_{v,k} \mu_k^2 {\boldsymbol \lambda}_k^T {\bf I}_M.
\end{eqnarray*}
Finally, the first iteration ends with node $N$ and the final update is given by
\begin{equation*}
{\mathds E} \left[ \left\| {\bf \overline w}_N (1) \right\|^2_{{\boldsymbol \sigma}_N} \right] = \left\| {\bf w}_o \right\|^2_{{\bf \overline F}_N {\boldsymbol \sigma}_N} + {\bf B}_N {\boldsymbol \sigma}_N.
\end{equation*}

Moving to iteration $2$, the update for node $1$ is given by
\begin{eqnarray*}
{\mathds E} \left[ \left\| {\bf \overline w}_1 (2) \right\|^2_{{\boldsymbol \sigma}_1} \right] &=& {\mathds E} \left[ \left\| {\bf \overline w}_N (1) \right\|^2_{{\bf F}_1 {\boldsymbol \sigma}_1} \right] + \sigma^2_{v,1} \mu_1^2 {\boldsymbol \lambda}_1^T {\boldsymbol \sigma}_1 \\
&=& \left\| {\bf w}_o \right\|^2_{{\bf \overline F}_N {\bf F}_1 {\boldsymbol \sigma}_1} + {\bf B}_N {\bf F}_1 {\boldsymbol \sigma}_1 + \sigma^2_{v,1} \mu_1^2 {\boldsymbol \lambda}_1^T {\boldsymbol \sigma}_1.
\end{eqnarray*}
For node $2$, we have
\begin{eqnarray*}
{\mathds E} \left[ \left\| {\bf \overline w}_2 (2) \right\|^2_{{\boldsymbol \sigma}_2} \right] &=& {\mathds E} \left[ \left\| {\bf \overline w}_1 (2) \right\|^2_{{\bf F}_2 {\boldsymbol \sigma}_2} \right] + \sigma^2_{v,2} \mu_2^2 {\boldsymbol \lambda}_2^T {\boldsymbol \sigma}_2 \\
&=& \left\| {\bf w}_o \right\|^2_{{\bf \overline F}_N {\bf \overline F}_2 {\boldsymbol \sigma}_2} + \left[ {\bf B}_N {\bf \overline F}_2 + {\bf B}_2 \right] {\boldsymbol \sigma}_2.
\end{eqnarray*}
Continuing for node $k$, we get
\begin{equation*}
{\mathds E} \left[ \left\| {\bf \overline w}_k (2) \right\|^2_{{\boldsymbol \sigma}_k} \right] = \left\| {\bf w}_o \right\|^2_{{\bf \overline F}_N {\bf \overline F}_k {\boldsymbol \sigma}_k} + \left[ {\bf B}_N {\bf \overline F}_k + {\bf B}_k \right] {\boldsymbol \sigma}_k.
\end{equation*}
The final update for iteration $2$ is given by
\begin{equation*}
{\mathds E} \left[ \left\| {\bf \overline w}_N (2) \right\|^2_{{\boldsymbol \sigma}_N} \right] = \left\| {\bf w}_o \right\|^2_{{\bf \overline F}_N^2 {\boldsymbol \sigma}_N} + {\bf B}_N \left[ {\bf \overline F}_N + {\bf I}_M \right] {\boldsymbol \sigma}_N.
\end{equation*}

Before moving on, we need to link the first update for node $k$ with the second update. Beginning with node $1$,
\begin{eqnarray*}
{\mathds E} \left[ \left\| {\bf \overline w}_1 (2) \right\|^2_{{\boldsymbol \sigma}_1} \right] &=& {\mathds E} \left[ \left\| {\bf \overline w}_N (1) \right\|^2_{{\bf F}_1 {\boldsymbol \sigma}_1} \right] + \sigma^2_{v,1} \mu_1^2 {\boldsymbol \lambda}_1^T {\boldsymbol \sigma}_1 \\
&=& {\mathds E} \left[ \left\| {\bf \overline w}_{N-1} (1) \right\|^2_{{\bf F}_N {\bf F}_1 {\boldsymbol \sigma}_1} \right] + \sigma^2_{v,1} \mu_1^2 {\boldsymbol \lambda}_1^T {\boldsymbol \sigma}_1 \\
&&+ \sigma^2_{v,N} \mu_N^2 {\boldsymbol \lambda}_N^T {\bf F}_1 {\boldsymbol \sigma}_1 \\
&=& {\mathds E} \left[ \left\| {\bf \overline w}_{N-2} (1) \right\|^2_{{\bf F}_{N-1} {\bf F}_N {\bf F}_1 {\boldsymbol \sigma}_1} \right] + \sigma^2_{v,1} \mu_1^2 {\boldsymbol \lambda}_1^T {\boldsymbol \sigma}_1 \\
&&+ \sigma^2_{v,N} \mu_N^2 {\boldsymbol \lambda}_N^T {\bf F}_1 {\boldsymbol \sigma}_1 \\
&&+ \sigma^2_{v,N-1} \mu_{N-1}^2 {\boldsymbol \lambda}_{N-1}^T {\bf F}_N {\bf F}_1 {\boldsymbol \sigma}_1 \\
&=& {\mathds E} \left[ \left\| {\bf \overline w}_1 (1) \right\|^2_{{\bf F}_2 \hdots {\bf F}_N {\bf F}_1 {\boldsymbol \sigma}_1} \right] + \sigma^2_{v,1} \mu_1^2 {\boldsymbol \lambda}_1^T {\boldsymbol \sigma}_1 \\
&&+ \sigma^2_{v,N} \mu_N^2 {\boldsymbol \lambda}_N^T {\bf F}_1 {\boldsymbol \sigma}_1 + \hdots \\
&&+ \sigma^2_{v,2} \mu_{2}^2 {\boldsymbol \lambda}_{2}^T \left\{ \prod\limits_{m=3}^{N} { {\bf F}_m} \right\} {\bf F}_1 {\boldsymbol \sigma}_1 \\
&=& \left\| {\bf w}_o \right\|^2_{ {\bf \overline F}_N {\bf F}_1 {\boldsymbol \sigma}_1 } + \left( {\bf B}_N {\bf F}_1 + \sigma_{v,1}^2 \mu_1^2 {\boldsymbol \lambda}_1^T {\bf I}_M \right) {\boldsymbol \sigma}_1
\end{eqnarray*}

Similarly, for node $2$, we get, after simplification
\begin{equation*}
{\mathds E} \left[ \left\| {\bf \overline w}_2 (2) \right\|^2_{{\boldsymbol \sigma}_2} \right] = \left\| {\bf w}_o \right\|^2_{ {\bf \overline F}_N {\bf \overline F}_2 {\boldsymbol \sigma}_2 } + \left( {\bf B}_N {\bf \overline F}_2 + {\bf B}_2 \right) {\boldsymbol \sigma}_2
\end{equation*}

Generalizing for node $k$ gives
\begin{equation*}
{\mathds E} \left[ \left\| {\bf \overline w}_k (2) \right\|^2_{{\boldsymbol \sigma}_k} \right] = \left\| {\bf w}_o \right\|^2_{ {\bf \overline F}_N {\bf \overline F}_k {\boldsymbol \sigma}_k } + \left( {\bf B}_N {\bf \overline F}_k + {\bf B}_k \right) {\boldsymbol \sigma}_k
\end{equation*}

The final update for iteration $2$ is thus give by
\begin{equation*}
{\mathds E} \left[ \left\| {\bf \overline w}_N (2) \right\|^2_{{\boldsymbol \sigma}_N} \right] = \left\| {\bf w}_o \right\|^2_{ {\bf \overline F}_N^2 {\boldsymbol \sigma}_N } + {\bf B}_N \left( {\bf \overline F}_N + {\bf I}_M \right) {\boldsymbol \sigma}_N
\end{equation*}
Since the final update for each iteration is given by the update of node $N$, we focus on node $N$ only for now. Continuing for node $N$, the update for iteration $i$ is given by
\begin{eqnarray}\label{eq15}
\nonumber {\mathds E} \left[ \left\| {\bf \overline w}_N (i) \right\|^2_{{\boldsymbol \sigma}_N} \right] &=& \left\| {\bf w}_o \right\|^2_{ {\bf \overline F}_N^i {\boldsymbol \sigma}_N } \\
&&+ {\bf B}_N \left( \sum\limits_{m=1}^{i-1} {\bf \overline F}_N^m + {\bf I}_M \right) {\boldsymbol \sigma}_N.
\end{eqnarray}
Similarly, for iteration $(i+1)$, we have
\begin{eqnarray}\label{eq16}
\nonumber {\mathds E} \left[ \left\| {\bf \overline w}_N (i+1) \right\|^2_{{\boldsymbol \sigma}_N} \right] &=& \left\| {\bf w}_o \right\|^2_{ {\bf \overline F}_N^{i+1} {\boldsymbol \sigma}_N } \\
&&+ {\bf B}_N \left( \sum\limits_{m=1}^{i} {\bf \overline F}_N^m + {\bf I}_M \right) {\boldsymbol \sigma}_N.
\end{eqnarray}
Subtracting (\ref{eq15}) from (\ref{eq16}), rearranging and simplifying gives
\begin{eqnarray}\label{eq17}
\nonumber {\mathds E} \left[ \left\| {\bf \overline w}_N (i+1) \right\|^2_{{\boldsymbol \sigma}_N} \right] &=& {\mathds E} \left[ \left\| {\bf \overline w}_N (i) \right\|^2_{{\boldsymbol \sigma}_N} \right] \\
\nonumber &&+ \left\| {\bf w}_o \right\|^2_{ {\bf \overline F}_N^{i+1} {\boldsymbol \sigma}_N } - \left\| {\bf w}_o \right\|^2_{ {\bf \overline F}_N^{i} {\boldsymbol \sigma}_N } \\
\nonumber &&+ {\bf B}_N \left( \sum\limits_{m=1}^{i} {\bf \overline F}_N^m + {\bf I}_M \right) {\boldsymbol \sigma}_N \\
\nonumber &&- {\bf B}_N \left( \sum\limits_{m=1}^{i-1} {\bf \overline F}_N^m + {\bf I}_M \right) {\boldsymbol \sigma}_N \\
\nonumber &=& {\mathds E} \left[ \left\| {\bf \overline w}_N (i) \right\|^2_{{\boldsymbol \sigma}_N} \right] + {\bf B}_N {\bf \overline F}_N^i {\boldsymbol \sigma}_N \\
&&+ \left\| {\bf w}_o \right\|^2_{ {\bf \overline F}_N^{i} \left[ {\bf \overline F}_N - {\bf I}_M \right] {\boldsymbol \sigma}_N }
\end{eqnarray}
The term ${\bf F}_N^{i}$ can be written in an iterative way as follows:
\begin{equation}\label{eq18}
{\bf A}_{N,i} = {\bf \overline F}_N^{i} = {\bf A}_{N,i-1} {\bf \overline F}_N.
\end{equation}
Inserting (\ref{eq18}) into (\ref{eq17}) gives the final update recursion
\begin{eqnarray}\label{eq19}
\nonumber {\mathds E} \left[ \left\| {\bf \overline w}_N (i+1) \right\|^2_{{\boldsymbol \sigma}_N} \right] &=& {\mathds E} \left[ \left\| {\bf \overline w}_N (i) \right\|^2_{{\boldsymbol \sigma}_N} \right] + {\bf B}_N {\bf A}_{N,i} {\boldsymbol \sigma}_N \\
&&+ \left\| {\bf w}_o \right\|^2_{ {\bf A}_{N,i} \left[ {\bf \overline F}_N - {\bf I}_M \right] {\boldsymbol \sigma}_N }.
\end{eqnarray}

Taking the weighting matrix as ${\bf \Sigma} = {\rm {\bf I}}_M$ results in the mean-square-deviation (MSD) and ${\bf \Sigma} = {\bf \Lambda}$ gives the excess mean square error (EMSE).

\subsection{Steady-State Analysis}
\label{SSAn}

For steady-state, we take a look at the relation between the first and second updates for node $N$ again. We do this in an iterative way as follows:
\begin{eqnarray*}
{\mathds E} \left[ \left\| {\bf \overline w}_N (2) \right\|^2_{{\boldsymbol \sigma}_N} \right] &=& {\mathds E} \left[ \left\| {\bf \overline w}_{N-1} (2) \right\|^2_{{\bf F}_N {\boldsymbol \sigma}_N} \right] + \sigma^2_{v,N} \mu_N^2 {\boldsymbol \lambda}_N^T {\boldsymbol \sigma}_N \\
&=& {\mathds E} \left[ \left\| {\bf \overline w}_{N-2} (2) \right\|^2_{{\bf F}_{N-1} {\bf F}_N {\boldsymbol \sigma}_N} \right] + \sigma^2_{v,N} \mu_N^2 {\boldsymbol \lambda}_N^T {\boldsymbol \sigma}_N \\
&&+ \sigma^2_{v,N-1} \mu_{N-1}^2 {\boldsymbol \lambda}_{N-1}^T {\bf F}_N {\boldsymbol \sigma}_N \\
&=& {\mathds E} \left[ \left\| {\bf \overline w}_{N-3} (2) \right\|^2_{{\bf F}_{N-2} {\bf F}_{N-1} {\bf F}_N {\boldsymbol \sigma}_N} \right] + \sigma^2_{v,N} \mu_N^2 {\boldsymbol \lambda}_N^T {\boldsymbol \sigma}_N \\
&&+ \sigma^2_{v,N-1} \mu_{N-1}^2 {\boldsymbol \lambda}_{N-1}^T {\bf F}_N {\boldsymbol \sigma}_N \\
&&+ \sigma^2_{v,N-2} \mu_{N-2}^2 {\boldsymbol \lambda}_{N-2}^T {\bf F}_{N-1} {\bf F}_N {\boldsymbol \sigma}_N \\
&=& {\mathds E} \left[ \left\| {\bf \overline w}_N (1) \right\|^2_{{\bf {\overline F}}_N {\boldsymbol \sigma}_N} \right] + {\bf B}_N {\boldsymbol \sigma}_N,
\end{eqnarray*}
where the last step is obtained by further successive iterations and simplifying. Generalizing for iteration $(i+1)$ gives
\begin{equation}\label{eq20}
{\mathds E} \left[ \left\| {\bf \overline w}_N (i+1) \right\|^2_{{\boldsymbol \sigma}_N} \right] = {\mathds E} \left[ \left\| {\bf \overline w}_N (i) \right\|^2_{{\bf {\overline F}}_N {\boldsymbol \sigma}_N} \right] + {\bf B}_N {\boldsymbol \sigma}_N.
\end{equation}

At steady-state, (\ref{eq20}) becomes
\begin{equation}\label{eq21}
{\mathds E} \left[ \left\| {\bf \overline w}_N (\infty) \right\|^2_{{\boldsymbol \sigma}_N} \right] = {\mathds E} \left[ \left\| {\bf \overline w}_N (\infty) \right\|^2_{{\bf {\overline F}}_N {\boldsymbol \sigma}_N} \right] + {\bf B}_N {\boldsymbol \sigma}_N.
\end{equation}

Rearranging and simplifying (\ref{eq21}) gives the steady-state equation
\begin{equation}\label{eq22}
{\mathds E} \left[ \left\| {\bf \overline w}_N (\infty) \right\|^2_{{\boldsymbol \sigma}_N} \right] = {\bf B}_N \left[ {\bf I}_M - {\bf \overline F}_N \right]^{-1} {\boldsymbol \sigma}_N.
\end{equation}

Using ${\bf \Sigma} = {\rm {\bf I}}_M$ results in the steady-state mean-square-deviation (MSD) and ${\bf \Sigma} = {\bf \Lambda}$ gives the steady-state excess mean square error (EMSE).

\section{Results and Discussion}
\label{results}

\begin{figure}[h]
\centering{\includegraphics[width=75mm]{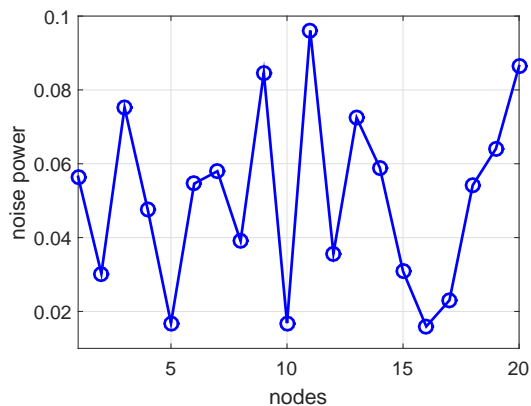}}
\caption{Noise power profile for each node for different SNR values.}\label{Fig2}
\end{figure}

This section compares the theoretical findings presented above with simulation results. A network of $N=20$ nodes is used for the simulations, with the length of the unknown vector being $M=4$. Results are shown for SNR values of $10$ dB, $20$ dB and $30$ dB. The noise power values at each node for the different SNR values are shown in Fig. \ref{Fig2}.

\begin{figure}[h]
\centering{\includegraphics[height=65mm]{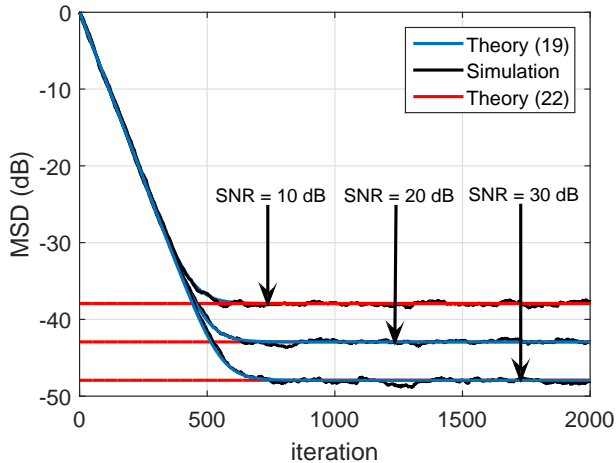}}
\caption{Theory (\ref{eq19}) and (\ref{eq22}) v simulation MSD comparison for white input data.}\label{Fig3}
\end{figure}

\begin{figure}[h]
\centering{\includegraphics[height=65mm]{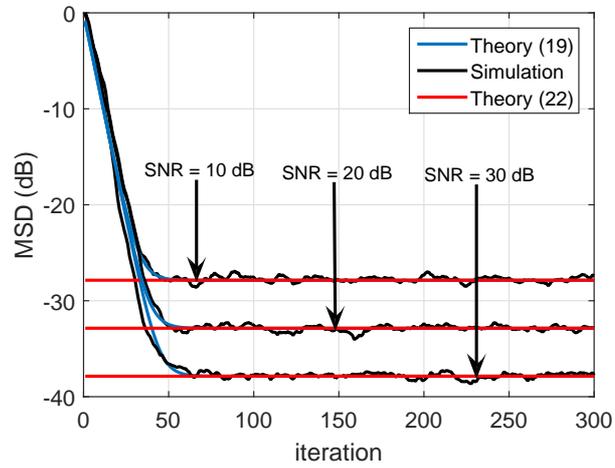}}
\caption{Theory (\ref{eq19}) and (\ref{eq22}) v simulation MSD comparison for white input data with fast convergence.}\label{Fig4}
\end{figure}

In the first experiment, the input data is assumed to be white. The convergence speed varies in order to have a comparison at both fast convergence and slow convergence. Therefore, the values used are $\mu=0.1$ and $\mu=0.01$. The results are shown in Figs. \ref{Fig3} and \ref{Fig4}. As can be seen, there is an excellent match between theory and simulation curves for all values of SNR. For comparison, the steady-state results obtained using (\ref{eq22}) are also shown.

\begin{figure}[h]
\centering{\includegraphics[height=65mm]{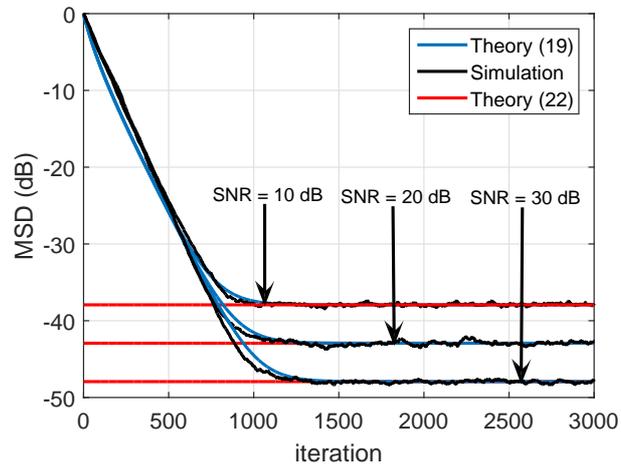}}
\caption{Theory (\ref{eq19}) and (\ref{eq22}) v simulation MSD comparison for correlated input data.}\label{Fig5}
\end{figure}

\begin{figure}[h]
\centering{\includegraphics[height=65mm]{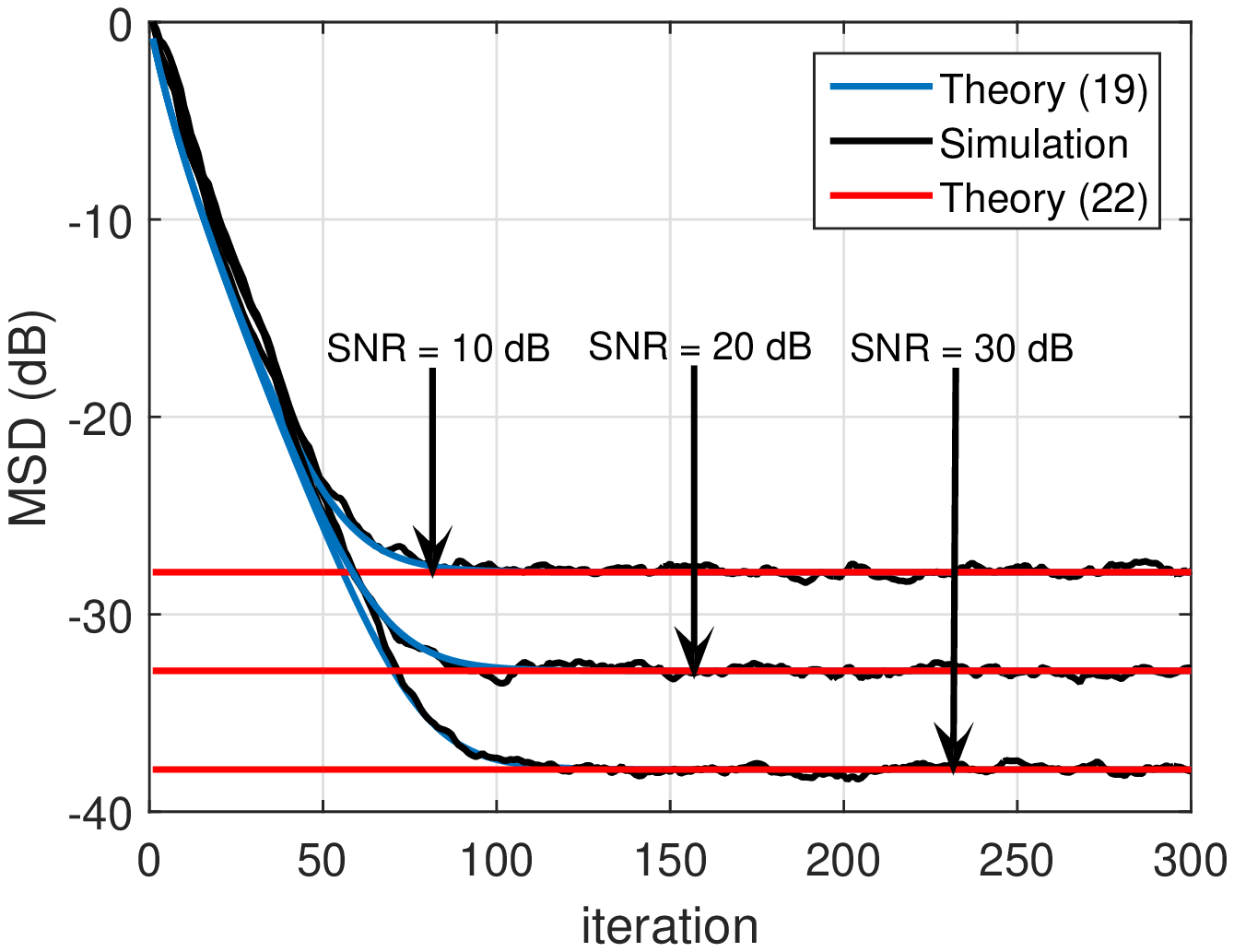}}
\caption{Theory (\ref{eq19}) and (\ref{eq22}) v simulation MSD comparison for correlated input data with fast convergence.}\label{Fig6}
\end{figure}

In the second experiment, the two scenarios are exactly the same except that the input data is now correlated, with the correlation factor being $0.4$. The results are shown in Figs. \ref{Fig5} and \ref{Fig6}. It should be noted that there is a slight discrepancy between the curves during the transient phase. However, this discrepancy is very low and the two curves are closely matched.


For further comparison, the steady-state results from the above experiments are listed below in Table \ref{Table1}. The aim of this table is to show the comparison between the results from (\ref{eq19}) and (\ref{eq22}). As can be seen, the results from both equations yield the same steady-state values.

\begin{table}[h]
\caption{Comparison of steady-state values under different scenarios.}
\label{Table1}
\begin{center}
\begin{tabular}{lccccc}
\hline\noalign{\smallskip}
\textbf{Data} & \textbf{SNR} & \textbf{Step-size} & \textbf{Sim.} & \textbf{Eq. (\ref{eq19})} & \textbf{Eq. (\ref{eq22})} \\
\noalign{\smallskip}\hline\noalign{\smallskip}
 & $10$ & $5e-3$ & $-37.96$ & $-37.93$ & $-37.93$ \\
 & & $5e-2$ & $-27.87$ & $-27.86$ & $-27.86$ \\
\textbf{White} & $20$ & $5e-3$ & $-42.95$ & $-42.93$ & $-42.93$ \\
 & & $5e-2$ & $-32.87$ & $-32.86$ & $-32.86$ \\
 & $30$ & $5e-3$ & $-47.97$ & $-47.93$ & $-47.93$ \\
 & & $5e-2$ & $-37.88$ & $-37.86$ & $-37.86$ \\
\noalign{\smallskip}\hline\noalign{\smallskip}
 & $10$ & $5e-3$ & $-37.93$ & $-37.93$ & $-37.93$ \\
 & & $5e-2$ & $-27.86$ & $-27.86$ & $-27.86$ \\
\textbf{Corr.} & $20$ & $5e-3$ & $-42.96$ & $-42.93$ & $-42.93$ \\
 & & $5e-2$ & $-32.85$ & $-32.86$ & $-32.86$ \\
 & $30$ & $5e-3$ & $-47.93$ & $-47.93$ & $-47.93$ \\
 & & $5e-2$ & $-37.86$ & $-37.86$ & $-37.86$ \\
\noalign{\smallskip}\hline
\end{tabular}
\end{center}
\end{table}

\section{Conclusion}
\label{conc}

This work presents the complete transient analysis for the incremental LMS algorithm, first presented in \cite{Lopes2007}. The learning behavior for the transient analysis is presented. Results show that the steady-state results from the transient analysis and the steady-state results from the steady-state analysis are an exact match. Furthermore, the simulation results closely match the theoretical results for different scenarios.

\end{document}